\documentstyle[revtexpl]{aps}
\begin{document}
\draft
\begin{title}
Proton halo of $^8$B in a microscopic model
\end{title}
\author{Attila Cs\'ot\'o\cite{email}}
\begin{instit}
Institute of Nuclear Research of the Hungarian Academy of Sciences \\
P.O.Box 51 Debrecen, H--4001, Hungary
\end{instit}
\receipt{29 June 1993}
\begin{abstract}
The experimentally predicted existence of a proton halo in $^8$B
is confirmed by dynamical microscopic multiconfiguration cluster
model calculations. The thickness of the proton halo in $^8$B is
0.5 $fm$ while in $^8$Li a 0.4 $fm$ thick neutron halo has been
found. It is shown that the huge quadrupole moment of $^8$B
partly comes from the very distortable $^7$Be core.
\end{abstract}

Recently the extensive experimental and theoretical study of
nuclei far from stability has revealed the neutron halo
structure of nuclei with large neutron excess
\cite{Tanihata1,Hansen}. A large variety of experiments have been
carried out \cite{Suzuki,Wiesbaden}, and numerous sophisticated
calculations have been performed
\cite{Danilin,Sagawa,Riisager1,Riisager2,Csoto} to understand
the physics
of these nuclei. One may think that the other limit of the
nuclear stability, the proton drip line, also floods us with a
mass of experimental and theoretical results on proton halo.

However, up till now there is only one experimental work which
claims the existence of a 0.8 $fm$ proton halo in $^8$B
\cite{haloprl} in order to explain the huge quadrupole moment
of this nucleus. Contrary to this finding, however, a
relativistic mean
field model of \cite{Tanihata2}, which describes well the thickness
of the neutron skin of various nuclei as a function of the
difference between the neutron and proton Fermi energies,
predicts that no proton halo exists in $^8$B, or if it exists
its thickness is certainly less than the thickness of the
neutron halo in $^8$Li and probably less then 0.2 $fm$. The
authors of \cite{Tanihata2,Tanihata} explained this result recalling that
because of the Coulomb barrier the proton could not move far
away from the $^7$Be core. But this argument overlooks the fact
that the addition of a Coulomb potential
raises the two-body potential not only in the barrier region but
also at the bottom of the potential, so the one proton
separation energy of $^8$B decreases comparing to $^8$Li with
almost
the same value as the height of the barrier. It means that the
height of the barrier which is seen by a neutron in $^8$Li is
roughly the same as seen by a proton in $^8$B. But while the
height of the barrier in $^7$Li+$n$ remains constant as $r$ goes
to infinity, in $^7$Be+$p$ it decreases by 1/$r$, a fact that
can causes the appearance of a proton halo. In contrast to what
was claimed in \cite{Tanihata2}, in \cite{Sagawa} a huge (0.9 $fm$)
proton halo was found in a shell model calculation. In addition
to this dispute, in \cite{Riisager2} the authors emphasized that
it is not certain that it is a proton halo which causes the
large quadrupole moment of $^8$B, it may partly originate from
the distortion of the core.

The aim of this Letter is to give a comprehensive microscopic
dynamical description of the structure of $^8$B and hereby
to make an attempt to settle the dispute related to the
existence of a proton halo in
this nucleus. For the sake of comparison we perform the
calculations for $^8$B and for its mirror partner $^8$Li
parallel.

Our model is of a resonating group type. We describe both $^8$Li and
$^8$B as three-cluster systems, i.e. $^8$Li=$\alpha+t+n$ and
$^8$B=$\alpha$+$h$(=$^3$He)+$p$. We take into account all
possible arrangements of these clusters and within each
arrangements we consider all angular momentum configurations of
any significance. For example the wave function of $^8$B in our
model is
\begin{eqnarray}
\Psi
&=&\sum_{S,l_1,l_2,L}\Psi_{S,(l_1l_2)L}^{p
(\alpha h)}+\sum_{S,l_1,l_2,L}\Psi_{S,(l_1l_2)L}^{h (\alpha
p)}+\sum_{S,l_1,l_2,L}\Psi_{S,(l_1l_2)L}^{\alpha (hp)}=
\nonumber \\
&=&\sum_{S,l_1,l_2,L}{\cal A}\left \{\left [ \left [\Phi
^p_i(\Phi ^\alpha\Phi^h)
\right ]_S
[\chi ^{\alpha h}_{l_1}(\mbox{\boldmath $\rho $}_{\alpha h})
\chi ^{p (\alpha h)}_{l_2}(\mbox{\boldmath $\rho $}_{p (\alpha h)})]
\raise-0.72ex\hbox{\scriptsize $L$}
\right ]_{JM}
\right \}
\nonumber \\
&+&\sum_{S,l_1,l_2,L}{\cal A}\left \{\left [ \left
[\Phi ^h(\Phi ^\alpha\Phi^p)\right ]_S
[\chi ^{\alpha p}_{l_1}(\mbox{\boldmath $\rho $}_{\alpha p})
\chi ^{h (\alpha p)}_{l_2}(\mbox{\boldmath $\rho $}_{h(\alpha
p)})] \raise-0.72ex\hbox{\scriptsize $L$}
\right ]_{JM}
\right \}
\nonumber \\
&+&\sum_{S,l_1,l_2,L}{\cal A}\left \{\left [ \left [\Phi
^\alpha(\Phi ^h\Phi^p)
\right ]_S
[\chi ^{h p}_{l_1}(\mbox{\boldmath $\rho $}_{h p})
\chi ^{\alpha (h p)}_{l_2}(\mbox{\boldmath $\rho $}_{\alpha (hp)})]
\raise-0.72ex\hbox{\scriptsize $L$}
\right ]_{JM}
\right \}.
\label{wfn}
\end{eqnarray}
\noindent
Here ${\cal A}$ is the intercluster antisymmetrizer,
$\Phi^\alpha$, $\Phi^h$, and $\Phi^p$ are the translation invariant
shell model states of the three clusters, the
\mbox{\boldmath $\rho $} vectors are the different intercluster
Jacobi coordinates, and [\ ] denotes angular momentum coupling.
Similar wave function was used in \cite{Baye} for $^8$Li except
that the $\alpha(tn)$ type components were omitted and the
two-cluster subsystems $(\alpha t)$ and $(\alpha n)$ were
described by pure two-center shell-model states. Putting
(\ref{wfn}) into the
eight-nucleon Schr\"odinger equation, we arrive at an equation for
the intercluster relative motion functions $\chi$. These
functions are expanded in terms of the so called tempered
Gaussian functions \cite{Kamimura} with different ranges and the
expansion coefficients are determined from a variational
principle. We have checked, by test calculations, that our basis
is spatially well-balanced and extensive enough. Further details
of the model can be found in \cite{Csoto}.

Building up the $2^+$ ground state of $^8$B and $^8$Li we
included all angular momentum components which can be physically
relevant. One may think that the inclusion of the $\alpha (hp)$
and $\alpha (tn)$
components is irrelevant because their thresholds are several MeV
above the ground state of $^8$B and $^8$Li. But it is not certain
{\em a priori} that such high lying channels cannot have an effect; e.g. it
turned out that in the ground state of $^6$He the effect of the
$t+t$ channel is significant \cite{Csoto}. In the $p(\alpha h)$ and
$n(\alpha t)$
components we included the $l_1=1$ and $3$ relative orbital angular
momenta which take place in the ${{3}\over{2}}^-$,
${{1}\over{2}}^-$, ${{7}\over{2}}^-$, and ${{5}\over{2}}^-$
states of $^7$Be and $^7$Li. In the $h(\alpha p)$ and $t(\alpha
n)$ components the $l_1=1$
partial wave which gives the
${{3}\over{2}}^-$ and ${{1}\over{2}}^-$ states of $^5$Li and
$^5$He was taken into account. And
finally in $\alpha (hp)$ and $\alpha (tn)$ the $l_1=1$ and $S=1$
values were
chosen in accordance with \cite{A=4} where the $^3$$P_1$ nature
of the low-lying $^4$Li and $^4$H states were claimed. In the $l_2$
angular momenta we restricted ourselves to $l_2\leq 3$. The angular
momentum space of our model in the $[S,(l_1l_2)L]J^\pi$ coupling
scheme is presented in table 1. We can say that in respect of the
relative motion space our model is close to the completeness.

The choice of the effective $N-N$ interaction is always a
crucial point of microscopic calculations. Here we follow the
philosophy of \cite{Csoto}, requiring that the description of
all subsystems appearing in the model should be as good as
possible. In the discussion of the proton halo of $^8$B all
models use $^7$Be+$p$ two-body picture
\cite{Sagawa,Riisager2,haloprl,Tanihata2}. We want to check the
validity of this assumption, therefore a reasonably good
description of $^7$Be and $^7$Li is indispensable.

To achieve this we chose the Minnesota force which had been
successfully used to describe the seven nucleon system
\cite{Tang}. Following \cite{Tang} we chose the common size
parameter of the various clusters in such a way that the sum of
the experimentally determined point matter rms radii of the
$\alpha$ particle and the triton is reproduced. A
common parameter $\beta =0.48$ $fm^{-2}$ was obtained. This
choice results in a bit too small radius for the $^3$He cluster
but it was found
\cite{Csoto} that the thickness of the halo is not too sensitive
to the absolute value of the radius within reasonable limits. If
we supplemented our
central interaction with the usual spin-orbit term of
\cite{Reichstein}, which e.g. gives reasonably good description
for $^6$He \cite{Csoto}, we found that it gives too strong
splitting between the ${{3}\over{2}}^-$ and ${{1}\over{2}}^-$
states of $^7$Be and $^7$Li. All spin-orbit terms of
\cite{Reichstein}
produced the same effect, that is why we changed the parameters
of the spin-orbit force \cite{Reichstein} and used the strength
$V_0=-25$ MeV and diffusity $d=1 fm$. The $u$ parameter of the
central interaction is set to
be $u=1.01$. As it can be expected, this spin-orbit force does
not give enough splitting between the ${{3}\over{2}}^-$ and
${{1}\over{2}}^-$ states of $^5$Li and $^5$He, but our tests
show that in our model it is impossible to get the correct
spin-orbit splittings for the seven and five nucleon systems with
the same spin-orbit force. Although not perfect, the
${{3}\over{2}}^-$ and ${{1}\over{2}}^-$ phase shifts of the five
nucleon systems are fairly good.

Some properties of $^7$Be and $^7$Li in our model are compared
in table 2 with experimental results. Note that the
${{7}\over{2}}^-$ and ${{5}\over{2}}^-$ states are unbound
in the $\alpha +h$ and $\alpha +t$ channel, respectively, that
is for them our model is a pseudobound
approximation. We also checked that the $\alpha +h$ and $\alpha
+t$ phase shifts are in good agreement with experiments. We can
say that the overall description of the seven nucleon systems
is reasonably good. We found that our $h+p$ and $t+n$ phase
shifts also agree well with experiments. This subsystem is the
only one in which the tensor force has a nonzero effect.
However, the rare experimental results \cite{li41,li42} are
contradicting concerning the low energy phase shift order of the
$^3$P$_0$ and $^3$P$_1$ states, which order would determine some
parameters of the tensor force. Because of the lack of these data even the
relation of the exchange parameters (i.e. the sign of the
strength) is uncertain. That is why we shall not use
tensor force in our calculations.

Using the full model space of table 1 we get 1.645 MeV
for the three-cluster separation energy in $^8$B (the
experimental value is 1.725 MeV) and 4.192 MeV in $^8$Li (the
experimental value is 4.501 MeV). Because the two-cluster
separation energies of the seven nucleon systems are correct we
get the same amounts of deviations from experiment for the
$^7$Be+$p$ and $^7$Li+$n$ separation energies as for the
three-cluster separation energies, above. These deviations
are not too large, and probably come from the not perfect
description of the five nucleon systems. The weights of the
various non-orthogonal channels (amount of clustering
\cite{amcl}) can be seen in table 1. The weights of the
different $SL$ components $(S,L)$=(1,1), (0,2), (1,2) are
96.98\%, 1.46\%, and 1.56\% in $^8$B and 96.11\%, 1.76\%, and
2.13\% in $^8$Li. To test the completeness of our basis in the
angular momentum space, we included the
$[S,(l_1l_2)L]=[1,(13)3]$ and, in the $p(\alpha h)$ and $n(\alpha
t)$ configurations, the $[1,(31)3]$ components and found that
their contribution to the binding energy is roughly 0.001 MeV,
their amount of clustering is around 0.01\% and the weight of
the $(S,L)$=(1,3) component is also roughly 0.01\%. These
figures confirm that our relative motion space is
nearly complete.

The quadrupole momenta and the various radii of $^8$B and $^8$Li
are shown in table 3, together with the results of a test run
we made to see the effect of the missing 0.3 MeV in $^8$Li. Note
that the experimental radii of \cite{Tanihata1} were revised for
certain nuclei \cite{Tanihata2}, and the new data analysis of
\cite{Tanihata2} would probably change the values of radii in
table 3. We
can see that in accordance with experiment the quadrupole moment
of $^8$B is much larger than that of $^8$Li. Actually our
quadrupole moment for $^8$Li seems to be too small, but we
remind of the fact that earlier experiments gave 2.4$\pm$0.2
$e$\hskip 0.03cm $fm^2$ for this quantity \cite{Ajzenberg}.
Our model predicts a 0.39 $fm$ thick neutron halo in $^8$Li,
while the thickness of the proton
halo in $^8$B is 0.52 $fm$. We tested the stability of these
figures by switching off some components in the wave function.
The stability of our results is extraordinary. For example if
we keep only the $\alpha (hp)$ channel and in this channel only
the $(S,L)=(1,1)$ component, although the three-cluster
separation energy of $^8$B decreases to 0.886 MeV the quadrupole
moment remains 5.78 $e$\hskip 0.03cm $fm^2$ and the thickness of the proton
halo is 0.46 $fm$.

We checked the validity of the $^7$Be+$p$ picture in three
calculations by keeping the three arrangements one by one in the
calculation. If the proton halo came from an arrangement
other than $p(\alpha h)$ this test would reveal it. But our
results are stable again, there is no essential difference
among the three arrangements, this is all the more conceivable
that, as it can
be seen in table 1, the overlaps of the $(S,L)=(1,1)$
components in the various arrangements are very large.

As far as the thickness of the proton halo is concerned, our
model predicts a smaller value comparing to \cite{haloprl} (0.8
$fm$) and to \cite{Sagawa} (0.9 $fm$), but this is certainly
caused by the fact that in these works the authors did not take
into account the distortion of the core. As we can see in table
2 the $^7$Be core is much more distortable than the
$^7$Li core, which means that the huge quadrupole moment of $^8$B
partly comes from the core deformations as was predicted in
\cite{Riisager2}. Our model, however, results in a much larger
proton halo than the upper limit of \cite{Tanihata2}, which
means that although the 0.4 $fm$ thick neutron halo of $^8$Li is
in
perfect agreement with \cite{Tanihata2}, the 0.5 $fm$ thick
proton halo in $^8$B does not fit to the strait line of
\cite{Tanihata2} on the $\Delta R^{rms} (\Delta E_F)$ plot.
The reason of this behaviour is not known to the author and
calls for an explanation.

In summary we have done careful microscopic multiconfiguration
multicluster calculations for $^8$Li and $^8$B. We have fixed
all parameters of our model to independent data, and the
description of all subsystems have been found reasonably good.
We have found a 0.4 $fm$ thick neutron halo in $^8$Li and a 0.5
$fm$ thick proton halo in $^8$B. We have pointed out that at
least a part of the
huge quadrupole moment of $^8$B is probably due to the fact
that the $^7$Be core is much more distortable than the $^7$Li
core. Our results are extraordinarily stable even against
radical changes of the model ingredients.

\mbox{}

This research was supported by OTKA (National Science Research
Foundation, Hungary) under contract numbers 3010 and F4348. The
author is indebted to Prof. B. Gyarmati for reading the manuscript.

\newpage
\begin{table}
\caption{Cluster decomposition of $^8$B and $^8$Li.}
\begin{tabular}{cccc}
\multicolumn{2}{c}{Clusterization}
                        & \multicolumn{2}{c}{Amount of clustering}\\
Partition & $S,(l_1l_2)L$ & $^8$B & $^8$Li \\
\tableline
$p(\alpha h)$, $n(\alpha t)$ & 1,(11)1 &0.9558&0.9465 \\
                             & 0,(11)2 &0.0131&0.0159 \\
                             & 1,(11)2 &0.0140&0.0193 \\
                             & 0,(13)2 &0.0003&0.0003 \\
                             & 1,(13)2 &0.0007&0.0009 \\
                             & 0,(31)2 &0.0126&0.0159 \\
                             & 1,(31)2 &0.0147&0.0202 \\
$h(\alpha p)$, $t(\alpha n)$ & 1,(11)1 &0.9582&0.9540 \\
                             & 0,(11)2 &0.0128&0.0157 \\
                             & 1,(11)2 &0.0137&0.0189 \\
                             & 0,(13)2 &0.0118&0.0151 \\
                             & 1,(13)2 &0.0137&0.0190 \\
$\alpha(hp)$, $\alpha(tn)$   & 1,(11)1 &0.9416&0.9363 \\
                             & 1,(11)2 &0.0127&0.0178 \\
                             & 1,(13)2 &0.0139&0.0192 \\
\end{tabular}
\label{tab1}
\end{table}

\begin{table}
\caption{Energies (relative to the $\alpha +h/t$ threshold),
quadrupole momenta and point nucleon rms radii of $^7$Be and
$^7$Li (the experimental energies are taken from
\cite{Ajzenberg}, the radii are extracted from \cite{Tanihata1}
by deconvoluting the finite nucleon size).}
\begin{tabular}{lr@{}lr@{}lr@{}lr@{}l}
&\multicolumn{4}{c}{$^7$Be}&\multicolumn{4}{c}{$^7$Li}
\\ \cline{2-5} \cline{6-9}
&\multicolumn{2}{c}{Theory}&\multicolumn{2}{c}{Experiment}&
\multicolumn{2}{c}{Theory}&\multicolumn{2}{c}{Experiment}\\
\hline
E(${{3}\over{2}}^-$) (MeV)& --1. & 59 & --1. & 59 & --2. & 47 & --2. & 47 \\
E(${{1}\over{2}}^-$) (MeV)& --1. & 14 & --1. & 16 & --1. & 99 & --1. & 99 \\
E(${{7}\over{2}}^-$) (MeV)&  3. & 07 &  2. & 98 &  2. & 37 &  2. & 16 \\
E(${{5}\over{2}}^-$) (MeV)&  3. & 31 &  5. & 14 &  2. & 87 &  4. & 21 \\
Q(${{3}\over{2}}^-$) ($e$\hskip 0.03cm $fm^2$)&  --6. & 35 & & &
                                 --3. & 76 &  --4.06&$\pm$0.08$^{\rm a}$ \\
                              &      &    & & &
                                     &    &  --3.7&$\pm$0.08$^{\rm b}$ \\
                              &      &    & & &
                                     &    &  --4.0&$\pm$1.1$^{\rm c}$ \\
$r_{\rm m}({{3}\over{2}}^-$) ($fm$)& 2.&43& 2.34&$\pm$0.03& 2.&39&
                                               2.36&$\pm$0.03 \\
                                 &   &  &   &           &   &  &
                                               2.39&$\pm$0.03$^{\rm d}$ \\
$r_{\rm n}({{3}\over{2}}^-)$ ($fm$)& 2.&37& 2.27&$\pm$0.03& 2.&44&
                                               2.41&$\pm$0.03 \\
$r_{\rm p}({{3}\over{2}}^-)$ ($fm$)& 2.&48& 2.39&$\pm$0.03& 2.&33&
                                               2.29&$\pm$0.03 \\
\end{tabular}
\tablenotes{$^{\rm a}$Ref.\ \cite{Ajzenberg}.}
\tablenotes{$^{\rm b}$Ref.\ \cite{Weller}.}
\tablenotes{$^{\rm c}$Ref.\ \cite{Vermmer}.}
\tablenotes{$^{\rm d}$Ref.\ \cite{Liatard}.}
\label{tab2}
\end{table}

\widetext
\begin{table}
\caption{Quadrupole momenta (in $e$\hskip 0.03cm $fm^2$), point nucleon rms
radii and the thickness of the halo (in $fm$) in the ground
state of $^8$B and $^8$Li.}
\begin{tabular}{lr@{}lr@{}lr@{}lr@{}lr@{}lr@{}lr@{}lr@{}lr@{}lr@{}l}
&\multicolumn{10}{c}{$^8$B}&\multicolumn{10}{c}{$^8$Li}
\\ \cline{2-11} \cline{12-21}
&\multicolumn{2}{c}{Q}&\multicolumn{2}{c}{$r_{\rm m}$}&
\multicolumn{2}{c}{$r_{\rm n}$}&\multicolumn{2}{c}{$r_{\rm p}$}&
\multicolumn{2}{c}{$r_{\rm p}-r_{\rm n}$}&
\multicolumn{2}{c}{Q}&\multicolumn{2}{c}{$r_{\rm m}$}&
\multicolumn{2}{c}{$r_{\rm n}$}&\multicolumn{2}{c}{$r_{\rm p}$}&
\multicolumn{2}{c}{$r_{\rm n}-r_{\rm p}$}\\
\hline
Tanihata \cite{Tanihata1} & & & & & & & & & & & & & 2.38&$\pm$0.03&
 2.44&$\pm$0.03 &2.27&$\pm$0.03 &0.&17 \\
Minamisono \cite{haloprl} &6.83&$\pm$0.21 & & &2.&20& 2.&98&
 0.&78& 3.27&$\pm$0.06&  &  & & & & & & \\
Kitagawa \cite{Sagawa} &7.&50& 2.&74& 2.&16& 3.&03& 0.&87 &
 3.&07& 2.&53& 2.&73& 2.&16& 0.&56 \\
Present &6.&58& 2.&57& 2.&25& 2.&74& 0.&52 &
 2.&25& 2.&45& 2.&59& 2.&20& 0.&39 \\
Present (fitted $^8$Li) & & & & & & & & & & &
 2.&21& 2.&42& 2.&56& 2.&18& 0.&38 \\
\end{tabular}
\label{tab3}
\end{table}


\begin{references}
\bibitem[*]{email} Electronic address: H988CSO@HUELLA.BITNET
\bibitem{Tanihata1} I. Tanihata et al., Phys. Rev. Lett. 55
(1985) 2676; I. Tanihata et al., Phys. Lett. 160B (1985) 380;
I. Tanihata et al., Phys. Lett. 206B (1988) 592.
\bibitem{Hansen} P.~G. Hansen and B. Jonson, Europhys. Lett.
4 (1987) 409; P.~G. Hansen, Nucl. Phys. A 553 (1993) 89c.
\bibitem{Suzuki} Proceedings of the International Symposium
on structure and reactions of unstable nuclei (Niigata, Japan,
1991), eds.\ K. Ikeda and Y. Suzuki (World Scientific,
Singapore, 1991).
\bibitem{Wiesbaden} I. Tanihata, Nucl. Phys. A 553 (1993) 361c;
T. Kobayashi, Nucl. Phys. A 553 (1993) 465c.
\bibitem{Danilin} B.~V. Danilin, M.~V. Zhukov, S.~N.
Ershov, F.~A. Gareev, R.~S. Kurmanov, J.~S. Vaagen and J.~M.
Bang, Phys. Rev. C 43 (1991) 2835; M.~V. Zhukov, B.~V. Danilin,
D.~V. Fedorov, J.~M. Bang, I.~J. Thompson and J.~S. Vaagen,
Submitted to Physics Reports 1992 [preprint NORDITA -- 92/90 N].
\bibitem{Sagawa} H. Kitagawa and H. Sagawa, Phys. Lett. B 299
(1993) 1.
\bibitem{Riisager1} A.~S. Jensen and Riisager, Nucl. Phys. A 537
(1992) 45; K. Riisager, A.~S. Jensen and P. M\o ller, Nucl. Phys.
A 548 (1993) 393.
\bibitem{Riisager2} K. Riisager and A.~S. Jensen, Phys. Lett. B
301 (1993) 6.
\bibitem{Csoto} A. Cs\'ot\'o, Phys. Rev. C, in press.
\bibitem{haloprl} T. Minamisono et al., Phys. Rev. Lett. 69
(1992) 2058.
\bibitem{Tanihata2} I. Tanihata et al., Phys. Lett. 289B (1992)
261.
\bibitem{Tanihata} I. Tanihata, in: Ref. \cite{Suzuki}, pp.233-240.
\bibitem{Baye} P. Descouvemont and D. Baye, Phys. Lett. B 292
(1992) 235.
\bibitem{Kamimura} M. Kamimura, Prog. Theor. Phys. Suppl. 68
(1980) 236.
\bibitem{A=4} D.~R. Tilley, H.~R. Weller and G.~M. Hale, Nucl.
Phys. A 541 (1992) 1.
\bibitem{Tang} Y. Fujiwara and Y.~C. Tang, Phys. Rev. C 28
(1983) 1869.
\bibitem{Reichstein} I. Reichstein and Y. C. Tang, Nucl. Phys.
A 158 (1970) 529.
\bibitem{Ajzenberg} F. Ajzenberg-Selove, Nucl. Phys. A 490
(1988) 1.
\bibitem{Weller} A. Weller et al., Phys. Rev. Lett. 55 (1985)
480.
\bibitem{Vermmer} W.~J. Vermmer et al., Phys. Lett. 138B (1984)
365.
\bibitem{Liatard} E. Liatard et al., Europhys. Lett. 13 (1990)
401.
\bibitem{li41} T.~A. Tombrello, Phys. Rev. 138 (1965) B40.
\bibitem{li42} J.~R. Morales, T.~A. Cahill, D.~J. Shadoon and H.
Willmes, Phys. Rev. C 11 (1975) 1905; G. Szaloky and F. Seiler
Nucl. Phys. A 303 (1978) 57.
\bibitem{amcl} R. Beck, F. Dickmann and R. G. Lovas, Ann. Phys.
(N.~Y.) 173 (1987) 1.
\end{references}
\end{document}